# Automating large-scale simulation and data analysis with OMNeT++: lessons learned and future perspectives

Antonio Virdis, Carlo Vallati, Giovanni Nardini
Dipartimento di Ingegneria dell'Informazione, University of Pisa
Largo Lucio Lazzarino 1, I-56122, Pisa, Italy
a.virdis@iet.unipi.it, carlo.vallati@iet.unipi.it, g.nardini@ing.unipi.it

*Abstract*—**Simulation is widely adopted in the study of modern computer networks. In this context, OMNeT++ provides a set of very effective tools that span from the definition of the network, to the automation of simulation execution and quick result representation. However, as network models become more and more complex to cope with the evolution of network systems, the amount of simulation factors, the number of simulated nodes and the size of results grow consequently, leading to simulations with larger scale. In this work, we perform a critical analysis of the tools provided by OMNeT++ in case of such large-scale simulations. We then propose a unified and flexible software architecture to support simulation automation.**

## I. Introduction

Nowadays simulation is a methodology widely used to drive the design and to assess performance of different computer systems. In computer networks in particular, simulation is widely adopted to drive the design of network or to assess the performance of existing deployments for provisioning or troubleshooting. Simulation models are exploited in place of real measurements or experiments for two main reasons: *(i)* simulation models can handle the complexity of such systems, characterized by many factors or settings that can influence the performance simultaneously; *(ii)* they can overcome the difficulty of studying systems that are distributed over distant areas and potentially all over the world.

In this context, OMNeT++ has gained popularity as a mature simulation tool. Especially in the area of networking, OMNeT++ is widely adopted by scientists and engineers that can exploit the availability of many simulation models for different network technologies, both wired and wireless.

Although simulation models are a simplified representation of actual systems, the increasing complexity of new communication technologies is currently pushing at a new different level the complexity of simulation models. Let us consider as example cellular networks: recent standards, e.g. LTE and LTE-Advanced, introduced new functionalities to handle the increasing demand for bandwidth and offer additional features to end users, with, however, a significant increase in complexity, which is necessarily reflected in the simulation models adopted, characterized by an overwhelming number of parameters, factors and number of simulated nodes.

Simulation models with a large number of factors and parameters usually imply simulation campaigns with a large number of different scenarios, aimed at evaluating the impact of each one on the overall system performance. Even though some techniques, e.g. factorial analysis [1], might be employed to reduce the number of scenarios, such simulation campaigns require a rigorous methodology to execute such large-scale experiments and, in particular, to analyze properly the large amount of results produced.

To this aim, software tools are usually employed to support the researcher to ensure a proper simulation workflow and eliminate - or minimize - biases or inaccuracies introduced by human operations, [2]. Specifically, tools that automate the execution of the simulation workflow and aid the researcher in the post-simulation analysis are usually employed. In this context, OMNeT++ already offers several tools and aids:

- An effective Graphical User Interface (GUI), which can be used to automate the execution of simulations. The end user can plan the simulation campaign through such interface exploiting an ad-hoc language adopted by OMNeT++ to configure the simulations and specify their parameters. The GUI can be used to run the experiments and monitor their progress through a graphical representation of network events.
- A post-simulation analysis GUI that can be exploited to visualize data and analyze metrics. Such GUI offers some basic data analysis operations, which can be exploited to produce simple graphs from simulation data.
- Some command line tools (opp_run) that drive and automate the execution of simulations without the GUI. Such tools can be used to run simulations on systems that lack of a graphical window system, e.g. a cluster or a server.
- A set of tools (scavetool) to export data from simulation results into different formats suited for external programs, e.g. Octave, Matlab, etc.

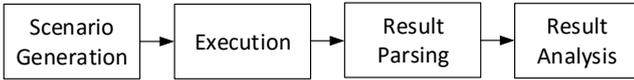

*Figure 1 - Main operations to be performed during a large-scale experiment.*

Although such functionalities are offered to automate the simulation workflow and aid researchers in post-simulation analysis, they have issues in handling large-scale simulation campaigns. Supporting complex simulations and analyzing large amount of results would require an improvement of the current software tools in order to ensure the rigorous execution of a consolidated simulation workflow. In this paper, we highlight the limits and issues of current tools included in the OMNeT++ suite in managing large-scale simulations and propose possible improvements in future perspective. Our goal is not to carry out a sterile analysis, but to trigger a fruitful discussion in the community on the best practices and their implementation, in order to improve future releases.

The remainder of the paper is organized as follows: in Section II simulation automation tools available for other network simulators are reviewed, in Section III the proper simulation workflow for large-scale simulations is presented along with an analysis of the automation tools currently available in OMNeT++. In Section IV we propose a software architecture for automating large-scale simulations, finally in Section V conclusions and future perspectives are presented.

## II. EXAMPLES OF SIMULATION AUTOMATION TOOLS

Network Simulator 2 (NS2) has been the standard de facto network simulator for years, before OMNeT++ and the release of its next major release NS3. Built around a simple basic architecture, NS2 became popular for the many network models made available over the years. Although popular, the simulator was completely lacking of a statistic collection framework and thus it offered no support for simulation automation or data analysis. To this aim, several third-party add-ons have been proposed over the year. Among them, it is worth to mention *ns2measure*, a framework designed to automate the collection of statistics, [3]. In addition, other extensions have been proposed to drive the simulation workflow and offer aid for post-simulation analysis. The *ANSWER* tool [2], for instance, is proposed to drive the experimental execution and automate the post-simulation data analysis through the aid of a GUI.

NS3, instead, provides native support for statistics collection and simulation flow management [7]. In particular, a set of internal modules are included to collect and store statistics, which can be saved on persistent storage (e.g. a database) or exported after simulation. Data provided by the *stat* module can be also exploited to check the status of simulations and drive their execution, e.g. stopping the simulation campaign when a certain level of confidence interval is reached. Although some functionalities for execution and data collection are offered, no support for post-simulation data analysis is provided natively as in OMNeT++.

## III. LARGE SCALE SIMULATION WORKFLOW

Consider as an example of large-scale simulation a network with hundreds of nodes of different types. Each type has a set of metrics, in the order of tens, that are measured over time for each node. The network and its nodes can be configured with hundreds of parameters to tune their behavior. Our goal is to perform a simulation campaign to assess the performance of the network in various configurations. For this purpose, we define as *fixed parameters* all the parameters that will assume the same value during the whole simulation campaign; we define, instead, as *factors* those parameters whose value will be varied during the campaign and that will actually modify the system configuration.

The common workflow is described in Figure 1: first, the set of simulation scenarios required to include all (or a subset) of the possible combination of parameters is generated, then all the simulations are executed and results collected and parsed, finally results are analyzed. In the following, we overview more in details each single phase.

### A. Scenario generation

The first step in the preparation of the simulation campaign is the generation of the scenario according to the parameters and factors. Within OMNeT++ this is done by means of *.ini* files, which can be modified in the IDE, either manually or through a form. Parameters are defined by simply assigning a value to them whereas factors are specified by means of the so called *iteration variables*, i.e. assigning an array of values to one single parameter. An example of the definition of one parameter and two factors is given in Figure 3. One of the most important factor is the number of *repetitions*, i.e. the number of times one configuration will be executed with different seeds for random number generation. Multiple repetitions are in fact used to perform independent replicas of the same scenario to increase the statistical soundness of the results, e.g., to improve confidence intervals.

Once all the above elements are defined, OMNeT++ will automatically translate the whole set of factors into a simulation campaign composed of $N$ runs, one for each possible configuration, as we represent in the left part of Figure 2. If we indicate with $|fact_i|$ the number of values that are defined for factor-$i$, the total number of runs will be equal to:

$$N = \left(\prod_i |fact_i|\right) \times R$$

where $R$ is the total number of repetitions. Each run is then associated with a unique numeric ID that will identify the single simulation all over the process. This approach leads to a run identification that is factor-agnostic, thus losing correlation between the run itself and the value of its factors. As we will see in the following, the latter is fundamental in the whole workflow as they define a run in a semantic way, representing how the system is configured. Note that the association between IDs and factors is preserved unless a factor is changed. It is quite common, for example, to change a factor during the

```
**.parameter = 50
**.factA = ${ 50 , 100 }
**.factB = ${ 1 , 2 }
```

*Figure 3 - Parameters and factors definition in .ini files*

| ID | factA | factB |
|----|-------|-------|
| 0  | 50    | 1     |
| 1  | 100   | 1     |
| 2  | 50    | 2     |
| 3  | 100   | 2     |

| ID | factA | factB | repetition |
|----|-------|-------|------------|
| 0  | 50    | 1     | 0          |
| 1  | 50    | 1     | 1          |
| 2  | 100   | 1     | 0          |
| 3  | 100   | 1     | 1          |
| 4  | 50    | 2     | 0          |
| 5  | 50    | 2     | 1          |
| 6  | 100   | 2     | 0          |
| 7  | 100   | 2     | 1          |

*Figure 2 - Example of mapping between run IDs and factors*

campaign, e.g. adding one factor (or a value), or performing additional repetitions of the same configuration in order to reach the desired statistical confidence. Every time a factor is added or more repetitions executed, the whole set of IDs is modified accordingly, as shown in the right part of Figure 2. This, however, modifies the correspondence between IDs and factors, which might lead to errors when results are analyzed.

### B. Multiple run support

Once the scenario is defined, the actual simulation campaign has to be performed, running the whole set of $N$ runs. Modern computers are equipped with multi-core processors, which can be exploited for the execution of multiple parallel simulations. Two main options are available in this respect.

The OMNeT++ environment offers a tool for running multiple batch simulations within the same machine, the *opp_runall* tool. The latter can be executed either via IDE or via command line and can be configured in terms of various parameters, e.g. the set of runs to be executed and the number of parallel processes. However, the simulations to be run are specified through their IDs, and the corresponding set of factors has to be retrieved manually. Moreover, opp_runall can be used only to execute one configuration file at a time, thus parallel execution of multiple configurations has to be performed running two different instances.

The second tool that is available in this respect is *AKAROA* [4]. The latter is a powerful framework for parallel execution of multiple simulations in different computers. It also offers the possibility to monitor at run-time a set of metrics, and extend the duration until some defined criteria are met. Although extremely powerful, AKAROA has a non-negligible setup cost, as it needs to be integrated within the simulator code, e.g. modifying the statistic collection. A few works on the integration of AKAROA within OMNeT++ are available in literature, e.g. [5]. However, considering the most recent research works and the activity within the community, it does not seem to be actually used.

### C. Post simulation parsing

After the whole campaign has successfully completed, results should be extracted and processed. One of the main problem with large-scale simulations is that they generate a considerable amount of result files, some of which can be very large. Parsing files can be cumbersome and also error prone.

The OMNeT++ environment has an extremely useful graphical tool for result extraction. First, it allows the selection of the set of files or folders to parse. Then, it has a powerful regular-expression based tool for parsing the results. The latter is extremely useful to quickly evaluate a small set of data. However, such tool does not scale with the size of results, i.e. it becomes extremely slow with large files and when the overall set of results becomes big. One common solution to the limits of the graphical interface in analyzing and extracting large volumes of simulation data is to exploit *scavetool*, a command line tool available to extract simulation data. Data, extracted in different format, can be imported in more powerful tools, e.g. Octave or Matlab, for analysis. However, *scavetool* has some limits, in particular when simulation scenarios with a large amount of data are considered, the tool is often unable to complete the extraction, as it requires all the data to be loaded in RAM. In addition, the extraction of a single metric or specific simulation scenarios is based on defining matching rules through regular expression, which is flexible but also error prone.

Another option available is to exploit R for data analysis. R is one of the most famous tool for statistical analysis. A plugin that allows to import directly in R simulation data is available. The researcher can first import data directly inside the R tool and then analyze the metrics and draw graphs. The main advantage of this approach is the large variety of statistical models and tools available in R, which makes possible the execution of any kind of analysis. The R tool, however, is not user-friendly and requires a non -negligible learning time. In addition, when very large simulation campaigns are considered, R cannot be used, as it requires all the data to be loaded in RAM.

When it comes to very large data sets, which cause memory issues to all the aforementioned solutions, the adoption of custom tools/scripts written by researchers is usually preferred, e.g. [8]. The realization of ad-hoc tools, however, is difficult, as simulation results are not stored in standard format, e.g. XML or JSON, and requires every time to re-invent the wheel.

### D. Results analysis

In the previous section, we mentioned that OMNeT++ IDE provides an efficient tool for quick evaluation of simulation results. Although this is very useful during the testing phase of a new models or algorithms, it is not sufficient to show results in a graceful and statistically sound way, i.e. for adequate presentation in a research paper. For example, it lacks of the possibility to evaluate confidence intervals for the mean values. Moreover, several types of chart are not available in the OMNeT++ environment, e.g. box plots and cumulative distribution functions (CDFs) plots, which are widely used in the research community for representing distributions. In some cases, more advanced analysis must be performed, e.g. factorial analysis. Thus, results must be processed by external tools such as Gnuplot, etc. However, such programs require to build custom scripts that operate on results (or a subset of it) provided in a predefined format, which could lead to the same issues highlighted for the development of custom tools.

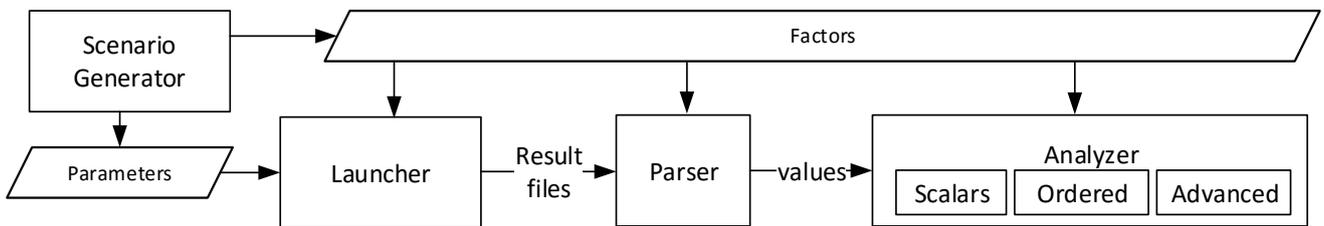

*Figure 4 - Proposed architecture for large-scale experiment*

## IV. A SOFTWARE ARCHITECURE FOR LARGE-SCALE SIMULATION

In this section, we will propose a software architecture for automating the execution of large-scale simulation campaigns. The purpose of this architecture is to serve as reference for future development, triggering a discussion within the community towards a consolidated point of view. We will take into account all the limitations highlighted in the previous sections and focus on four main goals:

1. define a modular and customizable structure;
2. use a factor-based indexing of configuration;
3. guarantee and improve statistical soundness of results;
4. ensure scalable performance.

We define four operational blocks, as represented in Figure 4, each one implementing one of the main steps of the simulation workflow described in Section III. The interactions between blocks occur using well-defined interfaces, but the internal structure of each of them can be customized.

Using Figure 4 as reference, the *scenario generator* creates two files: a first one containing all the fixed parameters of the campaign (the common .ini file generated through the OMNeT++ GUI); a second file, instead, containing all the factors with their values.

The *launcher* will take as input the parameters and the factors to execute the whole simulation campaign in parallel on a defined number of CPUs. The launcher should allow to execute selectively a subset of the simulation scenarios, e.g. for test or troubleshooting, with scenarios selected in a factor-based manner. This will allow one for example to execute all the campaigns where factor $x$ has value $y$. The number of repetitions will be also configurable, allowing dynamic extension.

The output of the *launcher* is a set of result files, each one tagged with the values of all the factors, which will be made available to the *parser*. The latter will translate the output of the simulator in the format expected by the *analyzer*. The *parser* can have various implementations depending on the format of the result files. For example if standard .sca files are used, it can be implemented as a wrapper for scavetool, still maintaining the aforementioned scalability issues, but with limited development cost. More efficient solutions can be obtained creating custom parses for the .sca files (e.g. [8]) or through the definition of a new format for data files from scratch (e.g. binary files) or adopting a standard format (e.g. XML, JSON, etc.).

Regardless of its internal implementation, the parser will produce a set of results, tagged with the values of the factors. The *analyzer* in turn will use such files to perform three main operations:

1. compute scalar results such as mean values;
2. create ordered statistics which can be used to generate CDFs, scatterplots, etc.;
3. perform advanced statistical processing, such as factorial analysis.

The analyzer has access to the list of factors; thus it can be configured to selectively operate on a subset of the results. Its final goal is to produce results that are ready for representation, thus any tool can be used to create plots, such as Excel, Kaleidagraph and Calc, or batch ones such as Gnuplot.

## V. CONCLUSIONS AND FUTURE PERSPECTIVES

In this work, we presented a critical analysis of the tools for simulation automation provided by the OMNeT++ framework. We focused on the context of large-scale simulations and discussed the limitation of such tool in each step of the simulation workflow. Finally, we proposed a software architecture for large-scale simulation, with the aim of serving as guideline for future development within the community. Our main goal is to trigger a discussion with all the members of the OMNeT++ community, and share our view on the subject.

## REFERENCES

[1] C. Cicconetti, E. Mingozzi, C.Vallati, "A 2k · r factorial analysis tool for ns2measure", In Proceedings of VALUETOOLS 2009.

[2] M. M. Andreozzi, G. Stea, C. Vallati, "A framework for large-scale simulations and output result analysis with ns-2", Simutools 2009.

[3] C. Cicconetti, E. Mingozzi, G. Stea. "An integrated framework for enabling effective data collection and statistical analysis with ns-2", WNS2 2006.

[4] K. Pawlikowski, V. W. C. Yau and D. McNickle, "Distributed stochastic discrete-event simulation in parallel time streams," Simulation Conference Proceedings, Winter, Lake Buena Vista, FL, USA, 1994.

[5] S. Sroka, H. Karl, "Using Akaroa2 with OMNeT++", 2nd OMNeT++ Workshop, Berlin, Germany, Jan 2002.

[6] http://www.r-project.org

[7] https://www.nsnam.org/docs/manual/html/statistics.html

[8] https://github.com/sommer/inet-sommer/tree/analysis/etc